\documentclass[12pt]{article}

\newcommand{\x}{arXiv:}
\newcommand{\m}{\mathrm}

\usepackage{graphicx}

\begin{document}
\thispagestyle{empty}
\begin{center}

\null \vskip-1truecm \vskip2truecm

{\Large{\bf \textsf{Universality of the Holographic}}}

{\Large{\bf \textsf{Angular Momentum Cutoff}}}

{\large{\bf \textsf{}}}

{\large{\bf \textsf{}}}

\vskip1truecm

{\large \textsf{Brett McInnes
}}

\vskip1truecm

\textsf{\\ National
  University of Singapore}
  \vskip0.3truecm
\textsf{email: matmcinn@nus.edu.sg}\\

\end{center}
\vskip1truecm \centerline{\textsf{ABSTRACT}} \baselineskip=15pt

\medskip
The AdS/CFT dual description of a peripheral heavy ion collision involves an asymptotically AdS rotating black hole. The explicitly known black holes of this kind, with planar event horizon topology [the ``KMV$_0$" spacetimes], have been shown to be unstable when string-theoretic effects are taken into account. It has been argued that this implies a ``holographic" angular momentum cutoff for peripheral collisions at very high energies. However, the KMV$_0$ black hole corresponds to a specific velocity distribution in the aftermath of a peripheral collision, and this distribution is not realistic at all points of the interaction zone. It could therefore be argued that the angular momentum cutoff is an artefact of this particular choice of bulk geometry. We demonstrate that, on the contrary, a Quark-Gluon Plasma with \emph{any} physically reasonable internal velocity distribution corresponds to a black hole which is still unstable, in the same way as the KMV$_0$ spacetime. The angular momentum cutoff is therefore a universal prediction of the holographic description of these collisions.

\newpage
\addtocounter{section}{1}
\section* {\large{\textsf{1. Peripheral Heavy Ion Collisions and Planar Black Holes}}}
One of the tasks of the LHC experiments is to study the Quark-Gluon Plasma [QGP] formed in collisions of lead nuclei \cite{kn:schuk}\cite{kn:schuk2}. Much attention, both experimental and theoretical, has been focused on the elliptic flow associated with peripheral [non-central] collisions; but the energies in this experiment are so high that it has also been proposed \cite{kn:liang}\cite{kn:bec}\cite{kn:huang} that novel phenomena [involving quark polarization] may eventually be observable in connection with the internal motion of the QGP. This is not implausible, in view of the very large angular momenta generated by ultra-high-energy peripheral collisions, a significant proportion of which can be transferred to the plasma.

In holographic approaches [see for example \cite{kn:solana}\cite{kn:pedraza}] to the study of the QGP, one begins by identifying an appropriate thermal AdS black hole spacetime to play the role of the bulk. In order to use holography to study a QGP endowed with a non-trivial amount of angular momentum, it is natural to turn to Carter's AdS-Kerr black hole spacetime \cite{kn:carter}\cite{kn:hawrot}, which has topologically spherical event horizons and boundary spatial geometry: this was pioneered by Atmaja and Schalm \cite{kn:schalm}. The idea is essentially that frame-dragging around a rotating black hole ---$\,$ an effect which persists at infinity for AdS black holes ---$\,$ can be used to model the shearing within the QGP.

In applying ideas from the AdS/CFT correspondence to the QGP or to condensed matter systems, the ultimate hope is that the gravitational dual theory will be able to reveal previously unknown properties of these very unusual forms of matter. The case of non-trivial internal angular momentum is particularly interesting from this point of view, because spinning black holes differ in an essential way from their static counterparts: angular momentum is a new ADM ``charge", and the AdS-Kerr metric cannot be obtained in any simple way from its AdS-Schwarzschild counterpart. In this sense, the introduction of angular momentum is a more drastic change on the bulk side of the correspondence than it appears to be on the boundary side, so we might hope that the former can indeed reveal new aspects of the latter.

In this spirit, it was proposed in \cite{kn:75} that there may be an \emph{upper bound on the ratio of the angular momentum density to the energy density} of the QGP produced in peripheral collisions. This arises in the following manner. The spherical spatial geometry of the conformal boundary of the AdS-Kerr black hole used in \cite{kn:schalm} clearly does not correctly represent the geometry in which the actual QGP is studied, which is of course flat. In the non-rotating case, one remedies this by using AdS black holes with event horizons having the topology of the ordinary Euclidean plane; as Witten observes in \cite{kn:confined}, one can obtain these by taking the infinite-mass limit of the spherical case. The rotating versions of these \emph{planar black holes} were discovered by Klemm, Moretti, and Vanzo \cite{kn:klemm}. Unlike their topologically spherical counterparts \cite{kn:74}, these are actually \emph{unstable} when string-theoretic effects in the bulk are included. This follows from an application of Seiberg and Witten's study \cite{kn:seiberg} of the stability of branes propagating in deformed versions of AdS. Assuming that the AdS/CFT duality holds here, we conclude that the ``spinning QGP" must likewise be unstable.

It was argued in \cite{kn:75} that one should expect this effect to be directly relevant, if at all, only to collisions at extreme energies and with favourable impact parameters, and so we interpret our result as imposing a ``cutoff" on the ratio of the angular momentum density to the energy density of the QGP. It is not entirely clear whether it is possible in practice to identify an experimental signature of such a cutoff in the case of the QGP. [One might expect, however, that it will play a role when the effects of vorticity on computations of the elliptic flow parameter v$_2$ are fully taken into account in simulation studies \cite{kn:bemo}.] The effect may in fact be equally important for the analogous condensed-matter \cite{kn:subir} systems [the ``rotating holographic superconductors" \cite{kn:sonner} and related systems \cite{kn:leigh}\cite{kn:leigh2}], and our methods can be adapted to apply to that case also. For simplicity, however, we shall formulate all of our results\footnote{Throughout this work, the bulk spacetime contains no matter other than a negative cosmological constant. For more general [but non-rotating] cases, see \cite{kn:AdSRN}\cite{kn:triple}\cite{kn:73}\cite{kn:ong1}\cite{kn:ong2}.} in the form in which they apply to the QGP, as it is produced in peripheral heavy-ion collisions.

However, before making any predictions regarding an angular momentum cutoff in real physical systems, one must confront the following important objection. The use of a planar geometry at infinity is clearly preferable to using a curved space, but the KMV$_0$ black hole shares one defect with the spherical AdS black hole: while both do allow us to use frame-dragging to model the QGP internal velocity distribution, \emph{neither of them leads to a distribution which is realistic.} We suggested in \cite{kn:75} that, despite this, the qualitative [and perhaps, to some extent, even the quantitative] conclusions we drew from a study of the KMV$_0$ case could still be trusted. However, more evidence for this claim is clearly essential. In other words, we must answer the question: if a realistic velocity distribution could somehow be modelled by a [different] planar AdS black hole, could this black hole be stable against the Seiberg-Witten effect? If this were so, then there would be no holographic angular momentum cutoff.

The purpose of this work is to show that the answer to this question is in the negative: a rotating planar AdS black hole inducing a realistic velocity profile at infinity would be unstable, in precisely the same way as the KMV$_0$ black hole. [In fact, we conjecture that \emph{every} planar AdS black hole with non-zero angular momentum is unstable in that sense.] This surprisingly general conclusion follows from an application of certain powerful theorems in global differential geometry, combined with a local analysis of special cases.

Thus we have a remarkable application of the holographic approach: \emph{it firmly predicts the existence of some kind of instability for the quark-gluon plasma produced in sufficiently energetic peripheral heavy ion collisions}. The rate at which such an instability develops is of course a different matter, which we do not try to address here.

We begin with a more precise formulation of the problem.

\addtocounter{section}{1}
\section* {\large{\textsf{2. Using Frame-Dragging to Model the QGP Velocity Profile}}}
The arguments presented in \cite{kn:75} were based on a simple geometric model of the conditions obtaining in the aftermath of a peripheral collision. As is explained in \cite{kn:bec}, the non-uniformity of the distribution of nucleons in the transverse direction [see Figure 1] implies that the amounts of momentum being carried left and right at a given value of x, the transverse coordinate, are not equal, and that the disparity is an increasing function of x. This generates a \emph{non-trivial velocity profile} in the x direction, giving rise to a shearing effect, which is the manner in which the QGP takes up a significant fraction of the total angular momentum of the system. [In the diagram, the z-axis is defined as the axis along which the momenta cancel; only the parts of the nucleons corresponding to positive x have been shown.]

\begin{figure}[!h]
\centering
\includegraphics[width=1.5\textwidth]{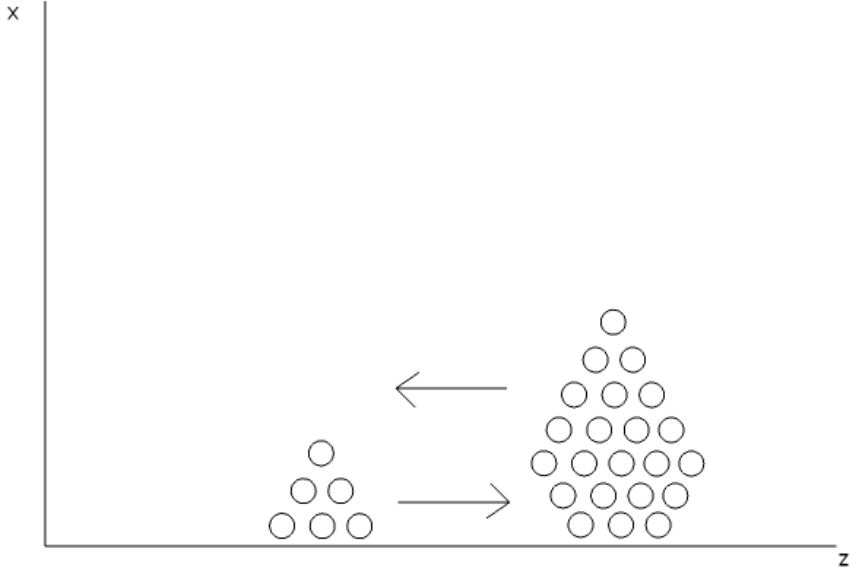}
\caption{Schematic Picture of Peripheral Collision [after \cite{kn:bec}].}
\end{figure}

The holographic dual of the field theory used to describe the ordinary QGP is an AdS-Schwarzschild spacetime [or an AdS-Reissner-Nordstr$\ddot{\m{o}}$m spacetime if one wants to have a non-zero chemical potential]. Hence it is very natural to model a QGP with a non-trivial angular momentum density by some kind of AdS-Kerr metric. The shearing we have just been describing will then correspond to \emph{frame-dragging} \cite{kn:visser} [or rather its counterpart at conformal infinity] around the black hole.

It is important to understand the nature of this correspondence. Frame-dragging is a purely geometric effect, like the expansion of the Universe in relativistic cosmology, and, as in that case, it is not subject to the familiar requirement that ``speeds" should not exceed the speed of light. However, if we wish to use it to \emph{model} the internal motion of the QGP, then of course we do have to ensure that this requirement is met. This means that we may have to impose conditions on the parameters of the black hole, not because they are required by black hole physics, but simply to ensure that the dual system respects causality.

A second point to bear in mind is that it is far from clear that frame-dragging can mimic the detailed velocity distribution that will result from the situation portrayed in Figure 1, even approximately. That distribution will be described by a profile [see \cite{kn:bec}\cite{kn:huang}] which vanishes along the z axis and then steadily increases, until it is bounded by causality. The magnitudes of the velocities will be determined by the density of the angular momentum, which of course can be quite large; in \cite{kn:75} it was measured by a parameter denoted a/L, where a [with units of length] is the ratio of the angular momentum and energy densities, and L [also with units of length] is related to the natural temperature scale of the QGP. We found that a/L could be quite large, perhaps even as large as 1000 at the LHC.

Let us examine these points with the aid of a concrete example, the AdS-Kerr black hole with a topologically spherical event horizon, as used in \cite{kn:schalm}. Then the geometry at infinity is described by a metric of the form
\begin{equation}\label{A}
\m{g(TSAdSK)_{\infty}\;=\;-\,dt^2 \;-\;{2\,a\;sin^2(\theta)\,dt d\phi \over 1 - (a/L)^2} \;+\; {L^2 d\theta^2 \over 1 - (a/L)^2cos^2(\theta)} \;+\; {L^2 sin^2(\theta)d\phi^2 \over 1 - (a/L)^2} },
\end{equation}
where $\theta$ and $\phi$ are the usual spherical angles on a topological two-sphere, a is the ratio of the ADM angular momentum of the black hole to its ADM mass,
and L is the asymptotic AdS curvature radius. Holographically, a/L corresponds to the QGP parameter discussed above.

This represents [a typical slice of] a system in which free particles with zero angular momentum, with $\theta$ = constant, rotate at a constant\footnote{Consider such a particle with a worldline having unit tangent $\m{\dot{t}\partial_t + \dot{\phi}\partial_{\phi}}$, where the dot denotes differentiation with respect to proper time along the worldline. Zero angular momentum implies that the inner product of this tangent vector with the Killing vector $\partial_{\phi}$ vanishes, and this yields $\dot{\phi}$ = $\m{a\,\dot{t}/L^2}$, giving the stated result.} angular velocity d$\phi$/dt = a/L$^2$ around the axis of symmetry. Such a particle is frame-dragged in the $\phi$ direction with a
linear ``velocity", taking into account the form of the metric in (\ref{A}), given by
\begin{equation}\label{AA}
\m{v \; = \; {(a/L)\,sin(\theta)\over \sqrt{1\,-\,(a^2/L^2)}}}.
\end{equation}
We see that this does depend on position in the ``transverse" direction, corresponding to the coordinate $\theta$. This is what we need, since the velocity distribution in the QGP has a non-trivial transverse profile. Causality in the model is respected provided that a/L satisfies a/L $<$ 1/$\sqrt{2}$ $\approx$ 0.7071, so that v $<$ 1 for all values of $\theta$.

The geometry described by the metric in equation (\ref{A}) has been used successfully to give an account of the drag force on a heavy quark moving through a QGP with non-zero angular momentum density \cite{kn:schalm}. However, there are several serious objections to it. First, the spatial geometry of the real system is of course planar, and not that of a deformed sphere; and, second, it is obvious that the velocity in the real system does not vary in the manner given by equation (\ref{AA}). Thirdly, the restriction a/L $<$ 1/$\sqrt{2}$ is far too strong, since, as we mentioned above, a/L in the real system is far larger.

In order to address these points, we proposed in \cite{kn:75} to study ``peripheral collision geometries" with metrics of the form
\begin{equation}\label{B}
\m{g_{PC} \;=\; - \, dt^2 \;-\; 2\omega_{\infty}(x \,, a/L)\, L \, dtdz \;+\; dx^2 \;+\; dz^2. }
\end{equation}
Here z is the axis of the collision, and x is the transverse coordinate, as above. The space described by these two coordinates, which take the place of $\phi$ and $\theta$ respectively in equation (\ref{A}), is now \emph{flat}. As before, the parameter a should be related to the ratio of the angular momentum and energy densities of the spinning QGP; L is a parameter with units of length, determined by a temperature scale characteristic of the QGP, and the dimensionless quantity $\omega_{\infty}$(x$\,$, a/L)L describes the transverse velocity distribution; note that this means that we must have
\begin{equation}\label{C}
\m{\omega_{\infty}(x\,, a/L)|_{x\, = \,0}\;=\;\omega_{\infty}(x\,, a/L)|_{a/L \,= \,0}\;=\;0.}
\end{equation}
The notation here expresses the idea that we want $\omega_{\infty}$(x$\,$, a/L) to have a dual interpretation as the asymptotic value of the angular velocity of some rotating AdS black hole [with ADM angular momentum related to the parameter a] with an event horizon necessarily having \emph{planar}, not spherical, topology\footnote{One easily verifies that free particles in the spacetime described by the metric in (\ref{B}), with x = constant and zero momentum in the z direction, do indeed satisfy dz/dt = $\omega_{\infty}$(x$\,$, a/L)L.}. That is, we would like to specify this function [by examining more-or-less realistic velocity profiles, such as those given in \cite{kn:bec} and \cite{kn:huang}], and then study an AdS planar black hole which induces at infinity the metric given in (\ref{B}). This black hole should allow large values of a/L. In this way we can hope to find an improved holographic description of the QGP when its angular momentum density is large.

The problem is of course to \emph{find} such a black hole. Indeed, even more basically, one would like to have some reassurance that such a black hole actually exists, for a prescribed function $\omega_{\infty}$(x$\,$, a/L). We shall discuss this question in the next Section; for the moment, let us remark that the familiar black hole uniqueness results do not apply directly to asymptotically AdS spacetimes, because one needs to prescribe much more data at infinity in this case; so one cannot easily rule out the possible existence of such objects.

\begin{figure}[!h]
\centering
\includegraphics[width=0.6\textwidth]{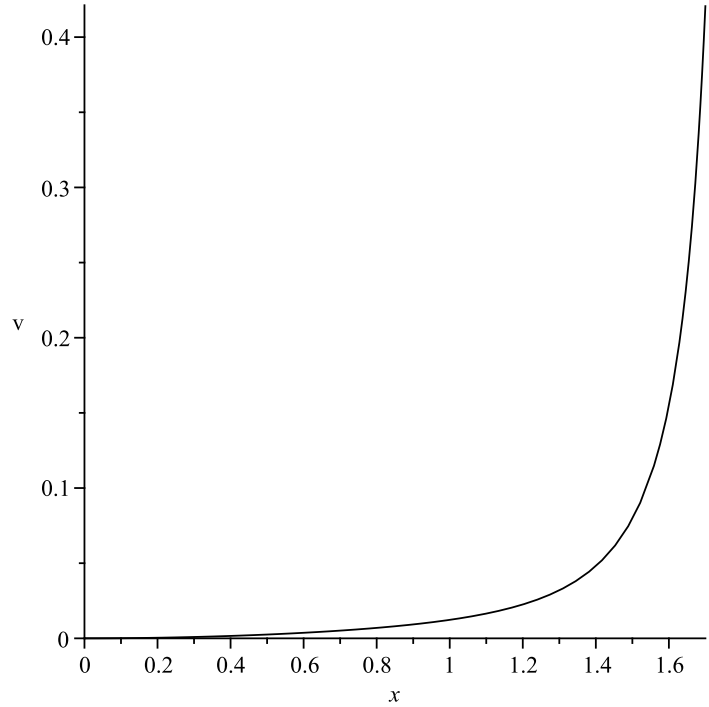}
\caption{Typical Velocity Profile Near Axis}
\end{figure}

In order to gain a first orientation, we proceeded in \cite{kn:75} in a different way: instead of trying to prescribe the geometry at infinity, we took the known rotating planar ``KMV$_0$" black hole \cite{kn:klemm}, and showed that it does indeed induce at infinity a metric of the form given in equation (\ref{B}); furthermore, it does allow [arbitrarily] large values of a/L. We then studied the corresponding $\omega_{\infty}$(x$\,$, a/L) function, and used the detailed structure of the bulk metric to show that the system is unstable in the Seiberg-Witten manner.

Now in fact it turns out that the KMV$_0$ velocity profile is very much like the profile used in \cite{kn:bec} [see the diagrams there and in \cite{kn:75}]. A profile of this sort is portrayed in Figure 2; this is in fact precisely a graph of the $\omega_{\infty}$(x$\,$, a/L) function for a KMV$_0$ metric, with particular values of a and L.

This velocity profile might be a reasonable approximation for nucleons \emph{near} to the axis, but it obviously fails globally, since it continues to increase indefinitely: ultimately it will violate causality. Indeed, in the very high-energy collisions with which we are concerned here, relativistic velocities would occur, post-collision, for values of x which are not very large relative to the size of the nucleus, and so the velocity profiles must \emph{flatten out} well before the boundary of the nucleus is approached. That is, a more realistic global velocity profile would resemble the one shown in Figure 3: crucially, there is an extended ``flat" region in which the velocities are [in natural units] close to unity. In fact, profiles of just this sort were used in the more recent work of Huang et al. \cite{kn:huang}. One can think of Figure 2 as an approximate representation of \emph{part} of the graph in Figure 3: the part which corresponds to values of x close to zero.

\begin{figure}[!h]
\centering
\includegraphics[width=0.6\textwidth]{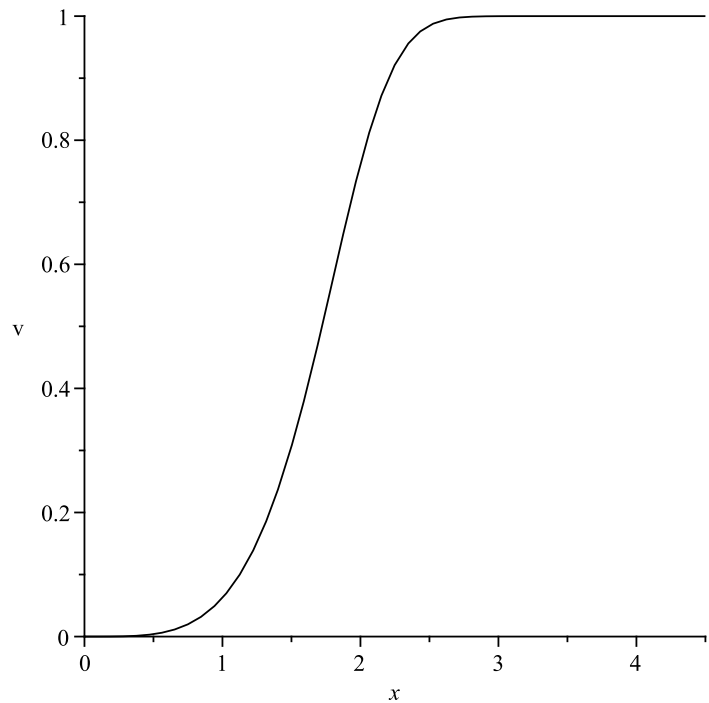}
\caption{Global Velocity Profile}
\end{figure}

Now the conclusions of \cite{kn:75} regarding the angular momentum cutoff were based on the use of the KMV$_0$ geometry, which means that we implicitly assumed a velocity profile of the same form as the one used in \cite{kn:bec}, that is, one of the sort shown in Figure 2. But, as we have just argued, this profile is only valid for a part of the interaction zone, not globally. The question stated in the Introduction can now be formulated more precisely: if the KMV$_0$ geometry could be deformed in such a way that the velocity at infinity is distributed as in Figure 3 instead of Figure 2, \emph{could that change the conclusion that the system must be unstable?} In other words, could it be that solving one problem [an unrealistic velocity profile for the QGP] might at the same time solve another [the instability of the dual black hole]?

This question can be answered, but some machinery must be developed first. We begin with a review of the KMV$_0$ spacetime and its possible deformed versions.

\addtocounter{section}{1}
\section* {\large{\textsf{3. The KMV$_0$ Geometry and its Deformations.}}}
One of the most basic features of asymptotically flat black holes is that they are [under well-understood conditions] uniquely specified by ADM data. This does not break down completely in the asymptotically AdS context, but the situation is certainly more subtle in this case.

The first and perhaps best-known novelty in the asymptotically AdS case is that
the event horizon need no longer have spherical topology \cite{kn:lemmo}. All of the following metrics represent (n+2)-dimensional AdS black holes of ADM mass M, with compact event horizons modelled on spaces of constant curvature k = $\{ -1, 0, +1 \}$ and asymptotic curvature $-1$/L$^2$:
\begin{eqnarray}\label{F}
\m{g(AdSBH^k_{n+2})} = -\, \m{\Bigg[{r^2\over L^2}\;+\;k\;-\;{16\pi M\over n V[W^k_n] r^{n-1}}\Bigg]dt^2\;} + \m{\;{dr^2\over {r^2\over L^2}\;+\;k\;-\;{16\pi M\over n V[W^k_n] r^{n-1}}} \;+\; r^2\,d\Omega^2[W_n^k].}
\end{eqnarray}
Here $\m{d\Omega^2[W_n^k]}$ is a metric of constant curvature k on an n-dimensional space W$\m{^k_n}$ with [dimensionless] volume $\m{V[W^k_n]}$. Thus, for a four-dimensional black hole with k = 0, the sections r = constant have the topology of a two-torus, with arbitrary volume. As is explained in more detail in \cite{kn:75}, we can ``decompactify" this example as follows. First, set M$^*$ = M/V[W$^0_2$]. Then the [two-dimensional] energy density of the black hole, evaluated at the event horizon, where r = r$_{\m{h}}$, is M$^*$/r$^2_{\m{h}}$. Now let both M and V[W$^0_2$] tend to infinity, in such a way that M$^*$ remains finite. Then we have a \emph{planar} black hole, described by a finite ``mass parameter" M$^*$:
\begin{eqnarray}\label{G}
\m{g(AdSBH^P_4)} = -\, \m{\Bigg[{r^2\over L^2}\;-\;{8\pi M^*\over r}\Bigg]dt^2\;} + \m{\;{dr^2\over {r^2\over L^2}\;-\;{8\pi M^*\over r}} \;+\;r^2\Big[d\psi^2\;+\;d\zeta^2\Big]},
\end{eqnarray}
where $\psi$ and $\zeta$ are dimensionless coordinates on the plane [so they have infinite ranges]. The quantity r$_{\m{h}}$ remains finite, and M$^*$/r$^2_{\m{h}}$ continues to have an interpretation in terms of an energy density. In this way we can avoid having a compact event horizon where necessary.

This spacetime has an obvious Euclidean version. Of course, the Euclidean version has a different geometry in the ``time" direction, and it also has a different \emph{topology} in that direction, since Euclidean ``time" is always compact. The local geometry in the spatial directions remains unchanged ---$\,$ in particular, it is still flat in the transverse directions parametrised by $\psi$ and $\zeta$ ---$\,$ but its topology can be chosen to be either planar or compact. Since the time coordinate is compactified in the Euclidean case, it seems natural to compactify in the spatial directions also, and, as we shall see later, this is still more compelling in the rotating case. If we do this, then the r = constant sections of the Euclidean black hole are [apart from the origin] copies of the three-torus T$^3$ [one circle corresponding to time, and one each to $\psi$ and $\zeta$]. The conformal structure at infinity here is clearly represented by a \emph{flat} metric on T$^3$.

The second new feature of the asymptotically AdS case is that these black holes can be \emph{deformed}, in the following sense. Take for example a five-dimensional static AdS black hole with a spherical event horizon. The spacetime geometry at infinity is that of the Einstein static universe, which has spatial sections described by the usual round metric on the three-sphere S$^3$. Now we deform this sphere to [say] a prolate spheroid, to obtain a ``squashed Einstein static universe":
\begin{equation}\label{PP}
\m{g(SqESU) = -\,dt^2\;+\;{L^2\over 4}\,\Big[\,(\sigma^1)^2\;+\;(\sigma^2)^2\;+\;s^2\,(\sigma^3)^2 \Big]}.
\end{equation}
Here $\sigma^{1,2,3}$ are orthogonal basis one-forms on the unit three-sphere S$^3$, L is the asymptotic AdS curvature radius, and s is a dimensionless constant non-zero ``squashing" parameter, which is greater than unity in the prolate case. We now ask whether there is a static black hole solution of the Einstein equation in the bulk [that is, with no matter other than a contribution due to the negative cosmological constant] which induces this prolate geometry at infinity. This question was answered numerically by Murata et al. \cite{kn:murata} [see also \cite{kn:kunz} and \cite{kn:brihaye}], who exhibited black hole metrics of precisely this kind:
\begin{eqnarray}\label{MUR}
\m{g^{MNT}} \;=\;\m{-\,F(r)e^{-2\delta (r)}\,dt^2 \;+\;{dr^2\over F(r)}}
\;+\;\m{{r^2\over 4}\,\Bigg[\,(\sigma^1)^2\;+\;(\sigma^2)^2\;+\;s(r)^2\,(\sigma^3)^2 \Bigg],}
\end{eqnarray}
where F(r) is a non-negative function, $\delta$(r) is a bounded function, vanishing at infinity, and s(r) is a function which can have an asymptotic positive value s$_{\infty}$ not necessarily equal to unity, so that indeed the geometry at infinity is given by the metric in (\ref{PP}).

One could follow an analogous procedure to obtain more complicated deformations of S$^3$, along several axes simultaneously \cite{kn:porrati}, and construct a similarly deformed black hole. This means that ``black hole uniqueness", in the stationary, asymptotically AdS context, has to involve a specification of the topology and geometry of the space at infinity. It also means that the AdS/CFT duality can potentially be applied to a very wide range of boundary geometries, just as was envisaged at the outset \cite{kn:wittenads}. This is crucial for many possible applications; see for example \cite{kn:kengo}.

Still more general deformations have been studied by Anderson et al. \cite{kn:chrusc1}\cite{kn:chrusc2}. They prove very remarkable results for the existence of distorted AdS black holes [both Lorentzian and Euclidean], of several topological types. Here we focus on the case of four-dimensional Euclidean AdS black holes with T$^3$ topology at infinity, of the kind discussed above. The main result of interest to us here runs as follows. Let M be the manifold D$^2$ $\times$ T$^2$, where D$^2$ is the two-dimensional disc [which ``fills in" some circle in the three-torus at infinity] and T$^2$ is the two-torus. Let $\mathcal{C}_s$(M) denote the space of conformal classes on the boundary represented by static [in the sense of Euclidean ``time"] metrics, and let $\mathcal{E}_s$(M) denote the inverse image of $\mathcal{C}_s$(M) under the usual bulk-to-boundary mapping of asymptotically hyperbolic Einstein bulk metrics to boundary conformal classes. [Note that $\mathcal{E}_s$(M) is represented by static bulk metrics.] Then Anderson et al. prove the following statement.

\bigskip
\noindent \textsf{THEOREM (Anderson-Chrusciel-Delay)}:
Each connected component of $\mathcal{E}_s$(M) containing a metric which is a Euclidean version of the metric given in equation (\ref{G}), with compactified (t, $\psi$, $\zeta$), is an infinite-dimensional space, with an image under the bulk-to-boundary map which contains open sets in $\mathcal{C}_s$(M).
\bigskip

In simple language: if one continuously deforms\footnote{Note that the deformations we are concerned with here, which always preserve a timelike Killing vector field, are quite different from those considered in \cite{kn:horror}.} the torus at infinity for the Euclidean k = 0 AdS black hole, so that it remains ``static" but ceases to be flat, then [unless the deformation eventually becomes so extreme that the conformal class of the torus leaves an open set in $\mathcal{C}_s$(M) around the original class] one can still find a ``static" Euclidean AdS black hole [having an Einstein metric of negative cosmological constant] which induces the deformed metric at infinity\footnote{This is of course strictly an existence proof. Attempts to exhibit such a black hole, even numerically, run into technical difficulties explained in \cite{kn:fragile}; deformed toral black holes with the \emph{simplest} structure [analogous to the one used in equation (\ref{MUR})] were shown not to exist.}. Analogous statements hold for Euclidean AdS black holes [see equation (\ref{F})] with k = $\pm$ 1, and also in the corresponding Lorentzian cases. This is the general result behind the particular constructions of \cite{kn:murata}\cite{kn:kunz}\cite{kn:brihaye}.

This discussion applies to static, that is, non-rotating black holes. However, Anderson et al. state that there is no obstacle in principle to extending their results to the stationary [rotating] case [though one should expect the technical problems to be substantial]. Considerable progress in this direction has been reported in \cite{kn:delay}. Again, one expects that bulk metrics will normally exist for physically reasonable deformations of the boundary. [Notice that giving angular momentum to an AdS black hole does have the effect of distorting the geometry at infinity; so, in a sense, stationary black holes are themselves a ``deformed" version of their static counterparts.]

What kinds of deformations might cause the existence theory to fail? Here is an example. Suppose that one takes the $\omega_{\infty}$(x$\,$, a/L) function corresponding to Figure 2, and replaces it by some function which is infinitely differentiable \emph{but not real-analytic}, that is, not everywhere given by its Taylor series. Then [\cite{kn:turk}, Theorem 2.1] the Euclidean version of the metric in equation (\ref{B}) is also infinitely differentiable but not necessarily real-analytic in harmonic coordinates; by adjusting $\omega_{\infty}$(x$\,$, a/L) suitably, one can ensure that it is not. The bulk Euclidean black hole metric would then also be infinitely differentiable but not analytic in harmonic coordinates, if it existed. But this is impossible, because the bulk metric is, by hypothesis, an Einstein metric, and it is known [\cite{kn:turk}, Theorem 5.2] that \emph{Einstein} metrics are always real-analytic in harmonic coordinates. Hence a ``deformation" of this kind would lead to a boundary metric with no bulk counterpart.

Mathematically, this situation arises because while the bulk has to have an Einstein metric, there is no such requirement for the boundary. Thus one should expect that there will be \emph{some} boundary deformations which cannot be matched in the bulk. However, the example we have given is rather exotic, and this suggests that these cases should be rare; normally we would not expect the distinction between analytic and infinitely differentiable metrics to be important for physical applications. Note, however, that the graph in Figure 3 does actually resemble the simplest examples of smooth but non-analytic real functions, namely those, like ``bump" functions, which are exactly constant on one part of their domain but not on another; and indeed we shall see an example of this sort, below. On the whole, however, we should not expect any difficulties as long as $\omega_{\infty}$(x$\,$, a/L) remains analytic.

In short, the metrics given by equation (\ref{F}) are merely \emph{representatives} of large classes of asymptotically AdS static black hole metrics, classes which are generated by continuously deforming the geometry at infinity of the representative spacetime; and one can expect a similar statement to hold in the rotating case.

For rotating black holes with topologically spherical event horizons, the basic representative is just Carter's AdS-Kerr metric \cite{kn:carter}\cite{kn:hawrot}. The planar analogue [generalizing the metric in equation (\ref{G}) above] is the KMV$_0$ metric of Klemm et al. \cite{kn:klemm}, to which we now turn.

We adopt the same [non-compact] coordinates as in equation (\ref{G}), (t, r, $\psi$, $\zeta$), and we define an angular momentum density parameter J$^*$ in a manner analogous to the definition of M$^*$, above. Set a = J$^*$/M$^*$; this will be interpreted holographically as the ratio of the angular momentum and energy densities in the boundary theory. [The angular momentum never occurs alone in the sequel, only in the form of this ratio.] Then the KMV$_0$ metric is given by
\begin{equation}\label{H}
\m{g(KMV_0) = - {\Delta_r\Delta_{\psi}\rho^2\over \Sigma^2}\,dt^2\;+\;{\rho^2 \over \Delta_r}dr^2\;+\;{\rho^2 \over \Delta_{\psi}}d\psi^2 \;+\;{\Sigma^2 \over \rho^2}\Bigg[\omega\,dt \; - \;d\zeta\Bigg]^2},
\end{equation}
where the asymptotic curvature is $-1$/L$^2$ and where
\begin{eqnarray}\label{I}
\rho^2& = & \m{r^2\;+\;a^2\psi^2} \nonumber\\
\m{\Delta_r} & = & \m{a^2+ {r^4\over L^2} - 8\pi M^* r}\nonumber\\
\Delta_{\psi}& = & \m{1 +{a^2 \psi^4\over L^2}}\nonumber\\
\Sigma^2 & = & \m{r^4\Delta_{\psi} - a^2\psi^4\Delta_r}\nonumber\\
\omega & = & \m{{\Delta_r\psi^2\,+\,r^2\Delta_{\psi}\over \Sigma^2}\,a}.
\end{eqnarray}
Notice that this metric reduces to the AdS planar black hole metric g(AdSBH$^{\m{P}}_4$) when a = 0; it is the planar analogue of Carter's topologically spherical AdS black hole, in the sense that it is an explicit metric with a Killing vector field [$\partial_{\zeta}$] in addition to the time translation field\footnote{As we shall see, it is analogous to Carter's metric in a much stronger sense: both induce conformally flat structures at infinity.}. We continue to refer to it as a ``planar" black hole, though it is planar only topologically, not geometrically. Its temperature and other details may be found in \cite{kn:klemm}\cite{kn:75}.

The KMV$_0$ metric induces at infinity a conformal structure represented by the metric
\begin{equation}\label{J}
\m{g(KMV_0)_{\infty}\; = \; -\,dt^2 \;-\;  2a\psi^2dtd\zeta \;+\; L^2\Big({d\psi^2\over 1 + a^2 \psi^4/L^2} +  d\zeta^2 \Big).}
\end{equation}
Setting dx = Ld$\psi$/$\m{\sqrt{1 + a^2 \psi^4/L^2}}$ and dz = Ld$\zeta$, we can write this as
\begin{equation}\label{K}
\m{g(KMV_0)_{\infty}\; = \: -\,dt^2 \;-\;  {2\over \gamma^2}\,sinl^2\Big(\sqrt{{a\over L}}\,\gamma\,{x\over L}\Big)dtdz \;+\; dx^2\; +  dz^2,}
\end{equation}
where i = $\sqrt{- 1}$, $\gamma$ = (1 + i)/$\sqrt{2}$, and sinl(x) $\equiv$ sn(x, i) is the ``lemniscatic sine", which will be described in detail in the next Section; it is one of the Jacobi elliptic functions \cite{kn:abramo}, sn(x, k), with imaginary elliptic modulus k = i. This function is real and bounded on the real axis; here, however, it is being evaluated along a diagonal axis in the complex plane. It is \emph{not} bounded in the complex plane; however, sinl($\gamma$x)/$\gamma$ is still real, as one can see from the series expansion of sn(x, k) [which can be shown \cite{kn:abramo} to take the form x + a$_5$x$^5$ + a$_9$x$^9$ + a$_{13}$x$^{13}$ ... for certain constant coefficients, when k = i.] This is analogous to the fact that sin(ix)/i = sinh(x) is real [and not bounded].

Comparing equation (\ref{K}) with equation (\ref{B}) we see that we have here a concrete example of the ``peripheral collision geometry": the two-dimensional space parametrised by x and z is flat, and the function $\omega_{\infty}$(x$\,$, a/L) in this case is just sinl$^2\Big(\m{\sqrt{{a\over L}}\,\gamma\,{x\over L}\Big)}/(\gamma^2$L). This is the function\footnote{In fact, the function sinl$^2\Big(\m{\sqrt{{a\over L}}\,\gamma\,{x\over L}\Big)}/\gamma^2$ is unbounded but periodic in the real variable x, because the Jacobi elliptic functions are doubly periodic in the complex plane. The periods in the real and imaginary directions coincide in this case, so that, with a = L = 1, the period of sinl$^2\Big(\m{\sqrt{{a\over L}}\,\gamma\,{x\over L}\Big)}/\gamma^2$ as a function of x is $\sqrt{2}\; \times$ 2K(i), where 2K(i) is the lemniscate constant discussed below. Figure 2 shows the function over [almost] one half of a period, between x = 0 and x = ($\sqrt{2}\; \times$ 2K(i))/2 $\approx$ 1.854, where it diverges.} graphed [with a = L = 1] in Figure 2; as explained earlier, it resembles the velocity distribution used in \cite{kn:bec}.

The task before us can now can be described concretely as follows. Begin with the metric in equation (\ref{K}): it describes a velocity distribution of the kind pictured in Figure 2. It arises holographically from the KMV$_0$ black hole, which has been shown \cite{kn:75} to be unstable to the effect discovered by Seiberg and Witten in \cite{kn:seiberg}. Suppose however that we continuously deform the boundary geometry, so that it continues to be of the ``peripheral collision" form, but with the graph in Figure 2 replaced by that in Figure 3. [This will also, of course, induce a continuous deformation of the Euclidean version of the spacetime; its geometry will change, but not its topology.] This deformation is actually very mild by comparison with the deformations contemplated in \cite{kn:chrusc1}: for in that work the authors were primarily interested in deformations which remove \emph{all} spacelike Killing vectors, whereas here we want deformations [see equation (\ref{B})] which retain $\partial_{\m{z}}$ as a Killing vector, both on the boundary and [therefore] in the bulk. That is, we do not want to lose any symmetry beyond that which has already been lost due to rotation; we only want to deform the azimuthal geometry of the black hole. In view of the above discussion, we shall assume henceforth that such black holes do exist.

The question is now: can it be that this deformed black hole \emph{is} stable, in the Seiberg-Witten sense? If so, this would nullify the conclusions of \cite{kn:75}: it would mean that we arrived at the conclusions of that work simply because we were using the ``wrong" black hole.

It may seem hopeless to try to answer such a general question, but fortunately there exist mathematical results of sufficient power to do so, except in certain special cases. We begin with a detailed study of the structure of the Euclidean version of the KMV$_0$ metric.

\addtocounter{section}{1}
\section* {\large{\textsf{4. The Topology of the KMV$_0$ Instanton}}}
Each KMV$_0$ metric defines a Euclidean instanton, given by the usual continuations of the time coordinate and of the parameter a:
\begin{equation}\label{L}
\m{g(EKMV_0) = {\Delta^E_r\Delta^E_{\psi}\rho_E^2\over \Sigma_E^2}\,dt^2\;+\;{\rho_E^2 \over \Delta^E_r}dr^2\;+\;{\rho_E^2 \over \Delta^E_{\psi}}d\psi^2 \;+\;{\Sigma_E^2 \over \rho_E^2}\Bigg[\omega_E\,dt \; - \;d\zeta\Bigg]^2},
\end{equation}
where we use the same names for Euclidean and Lorentzian coordinates, and where the component functions are straightforward continuations of the formulae (\ref{I}) above. It is convenient to introduce a new coordinate $\xi$, replacing $\psi$, defined\footnote{In the following discussion, we take a to be strictly non-zero; the non-rotating case has to be treated separately, as above, and not as the limit in which a tends to zero.} by
\begin{equation}\label{M}
\m{\psi \;=\; \sqrt{{L \over a}}\,sinl\Big(\sqrt{{a\over L}}\,\xi\Big)};
\end{equation}
here sinl(x), the ``lemniscatic sine" function mentioned earlier, is defined as sinl(x) $\equiv$ sn(x, i), one of the Jacobi elliptic functions \cite{kn:abramo}. Its graph, for real x, is shown in Figure 4.

\begin{figure}[!h]
\centering
\includegraphics[width=0.6\textwidth]{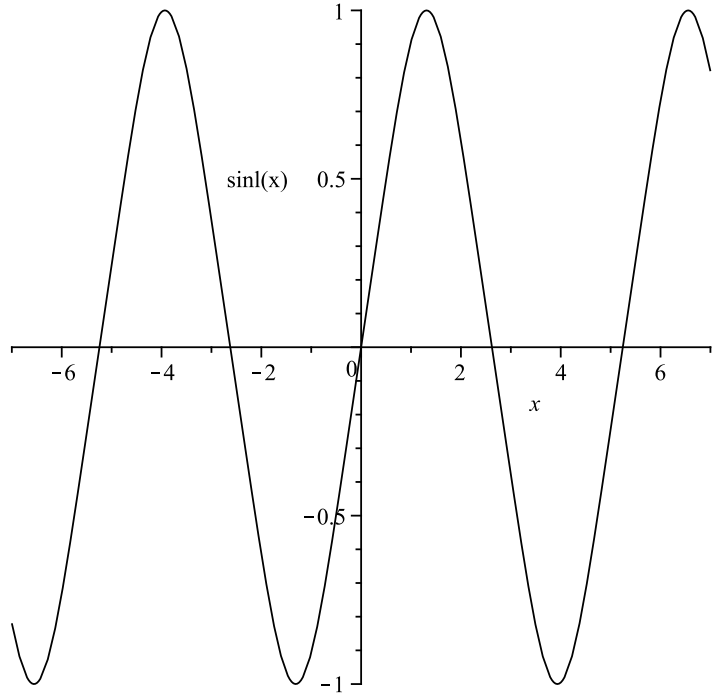}
\caption{Lemniscatic sine.}
\end{figure}

In terms of this coordinate, the metric is
\begin{equation}\label{N}
\m{g(EKMV_0) = {\Delta^E_r\rho_E^2\,cosl^2\Big(\sqrt{{a\over L}}\,\xi\Big)\over \Sigma_E^2}\,dt^2\;+\;{\rho_E^2 \over \Delta^E_r}\,dr^2\;+\;\rho_E^2\,d\xi^2 \;+\;{\Sigma_E^2 \over \rho_E^2}\Bigg[\omega_E\,dt \; - \;d\zeta\Bigg]^2},
\end{equation}
where
\begin{eqnarray}\label{O}
\m{\rho_E^2} & = & \m{r^2\;-\;aL\;sinl^2\Big(\sqrt{{a\over L}}\,\xi\Big)} \nonumber\\
\m{\Delta^E_r} & = & \m{- a^2+ {r^4\over L^2} - 8\pi M^*r}\nonumber\\
\m{\Sigma_E^2} & = & \m{r^4\,cosl^2\Big(\sqrt{{a\over L}}\,\xi\Big) + L^2\Delta^E_r\,sinl^4\Big(\sqrt{{a\over L}}\,\xi\Big)}\nonumber\\
\m{\omega_E} & = & \m{{{L\over a}\,\Delta^E_r\,sinl^2\Big(\sqrt{{a\over L}}\,\xi\Big)\,+\,r^2\,cosl^2\Big(\sqrt{{a\over L}}\,\xi\Big)\over \Sigma_E^2}\,a},
\end{eqnarray}
where cosl(x), the ``leminiscatic cosine", is defined as the derivative of sinl(x); it can be expressed in terms of the standard Jacobi elliptic functions \cite{kn:abramo} cn(x, k) and dn(x, k) as cosl(x) = cn(x, i)dn(x, i). Like sinl(x), cosl(x) is periodic and varies between $\pm$1 along the real axis; in fact, sinl$^4$(x) + cosl$^2$(x) = 1; see Figure 5.

\begin{figure}[!h]
\centering
\includegraphics[width=0.6\textwidth]{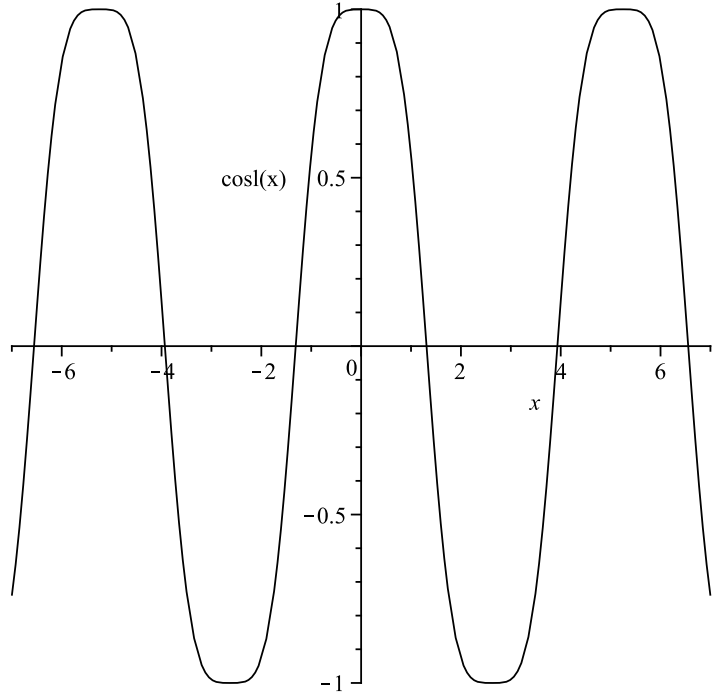}
\caption{Lemniscatic cosine.}
\end{figure}

It is of interest to examine the boundary metric: it is given by
\begin{equation}\label{P}
\m{g(EKMV_0)_{\infty}\; = \; d\tau^2 \;-\;  2\,sinl^2\Big(\sqrt{{a\over L}}\,\xi\Big)d\tau d\zeta \;+\; d\xi^2 \;+\;  d\zeta^2,}
\end{equation}
where we have set $\tau$ = t/L, and factored out a common L$^2$. As usual, the $\xi$-$\zeta$ plane is still flat. In the non-rotating case, we \emph{chose} the topology to be toral, but here we will see that this topology is the physical one in the rotating case. [Note that (\ref{P}) can be obtained directly from (\ref{K}) above by complexifying t and a.]

The point to note here is that the lemniscatic sine and cosine functions occurring in (\ref{N}) and (\ref{O}) are being evaluated along the \emph{real} axis, which was not the case in our earlier discussion of the Lorentzian boundary metric. In this case, they behave very much like the ordinary circular sine and cosine. The periods of sinl(x) and cosl(x) are given by 4K(i), where K(k) is the complete elliptic integral of the first kind \cite{kn:abramo}, which is real even when evaluated along the imaginary axis. Because these two functions occur in the formulae only in squared form, this means that every component of the metric in (\ref{N}) is periodic in $\xi$ with period 2K(i)$\sqrt{\m{L/a}}$. Unlike the function sinl($\gamma$x)/$\gamma$ which occurs in the Lorentzian case, however, all of these components are in addition \emph{bounded} everywhere, so that they are well-defined for all real $\xi$, as Figures 4 and 5 show.

In fact, sinl(x) $\equiv$ sn(x, i) is just Gauss' ``sinus lemniscaticus" \cite{kn:lemniscate}, the function which arises when one computes the arc length of a lemniscate. In polar coordinates, the unit lemniscate has the equation r$^2$ = cos(2$\theta$), so that [for the right lobe of the curve] $\theta$ ranges between $- \pi$/4 and $\pi$/4, both values corresponding to the  origin. If $\sigma$ denotes [signed] length along this lobe of the lemniscate, measured away from the point (r = 1, $\theta$ = 0), then sinl($\sigma$) = tan($\theta$), and we see that the point with $\sigma$ = K(i) [$\theta$ = $\pi$/4] is identified with the point having $\sigma$ = $-$K(i) [$\theta$ = $- \pi$/4], in precisely the same way as the points $\theta$ = $\pm \pi$ are identified on the unit circle.
[Here 2K(i) is a real number \cite{kn:lemnconst}, the ``lemniscate constant" $\approx$ 2.622058; it plays a role analogous to $\pi$ for the circle.] This means that the map $\sigma$ $\, \rightarrow \,$ 4$\,$arctan[sinl($\sigma$)] gives a homeomorphism from one lobe of the lemniscate to the circle. Thus our coordinate $\sqrt{\m{a/L}}\,\xi$, which corresponds to length along a lobe of a lemniscate, can also be regarded as a coordinate on a circle. In short, \emph{the geometry is compactified in the $\xi$ direction}, with a period which depends on a, the ratio of the angular momentum and mass parameters of the original black hole.

On the other hand, as with any black hole instanton, the ``time" coordinate here has to be periodic, with period given by the reciprocal of the Hawking temperature. Since d$\zeta$ is mixed with dt, this in turn means \cite{kn:hawrot}\cite{kn:75} that $\zeta$ is periodic [with a period which can be computed given the temperature and the parameter a]. In short, every non-radial coordinate in equation (\ref{N}) is periodic, though the three periods differ in general. Furthermore the volume form of the metric on the three-dimensional slices r = constant never vanishes\footnote{The relevant determinant is given by $\m{\Delta^E_r\,\rho_E^2}$. The proof that neither factor vanishes away from the origin is given in \cite{kn:75}.} except at the origin of the Euclidean space; this ensures that none of these slices is ``pinched off" at any point, in these coordinates. It follows that the r = constant sections, other than the origin, have the topology of a torus. \emph{This means that the boundary too is necessarily of toral topology in the Euclidean case.} This will be important in the sequel. [The \emph{geometry} of this torus is given by the metric in equation (\ref{P}): note that it is not flat.]

We may now turn to the question of the stability of the black holes obtained by deforming the KMV$_0$ spacetime in the manner we have described earlier.

\addtocounter{section}{1}
\section* {\large{\textsf{5. Seiberg-Witten Instability of Deformed KMV$_0$: the Generic Case}}}
In their study of deformations of the spherical AdS$_5$-Schwarzschild black hole, Murata et al. \cite{kn:murata} noticed the following curious fact. The scalar curvature of the metric given in equation (\ref{PP}), and also of its Euclidean version, is
\begin{equation}\label{PPP}
\m{R[g(SqESU)]\;=\;{2\over L^2}\,\Big[\,4\;-\;s^2 \Big].}
\end{equation}
We see that if the 3-sphere at infinity becomes sufficiently prolate [s $>$ 2], then the scalar curvature at infinity can become negative. This causes a tachyonic mode to arise in a scalar field on the boundary. Yet, as Murata et al. emphasise, the bulk black hole [with metric given in equation (\ref{MUR}), above], which has a rather mildly prolate event horizon but is otherwise apparently unremarkable, shows no obvious sign of being unstable. This is potentially a very important observation, because it constitutes a direct challenge to the validity of the AdS/CFT correspondence.

The resolution is as follows \cite{kn:fragile}. Seiberg and Witten \cite{kn:seiberg} observed that deformations of any bulk AdS-like geometry have important consequences for extended objects such as branes propagating in that bulk. In particular, the action for a BPS brane is a function of radial position, and they showed that this action will become [and remain] negative beyond some value of the radial coordinate if the Euclidean version of the bulk metric has \emph{negative scalar curvature} at infinity\footnote{The scalar curvature of a Riemannian metric on a compact manifold can always be made constant by means of a conformal transformation \cite{kn:schoenyam}, and it is to this constant that we refer when we speak of the sign of the scalar curvature at infinity.}. This signals an instability, since brane-antibrane pairs will nucleate, and a brane can lower its action by expanding instead of contracting. Thus, the distorted black hole with negative Euclidean scalar curvature at infinity studied by Murata et al. \emph{is} in fact unstable from a string-theoretic point of view\footnote{Similarly, black holes with compact hyperbolic event horizons \cite{kn:lemmo} are unstable in this way.}, and it becomes so precisely when the boundary scalar ceases to be stable. In this way, the AdS/CFT correspondence is preserved. We see that the Seiberg-Witten effect plays a crucial role in the correspondence, and its presence or absence must always be checked whenever an asymptotically AdS spacetime is deformed in any way.

Now let us try to apply this to the KMV$_0$ black hole. What is the scalar curvature of the metric in equation (\ref{P}), after it has been conformally transformed to a constant? The answer is that it is precisely zero, as we shall show in detail in the next Section. We see that, in this case, the criterion we have just been discussing fails us. We have to examine the brane action explicitly, and this was done in \cite{kn:75}. The result was that the brane action does become and remain negative beyond a certain value of the radial coordinate, so the KMV$_0$ black hole is indeed unstable. [This is of course not a contradiction: the criterion given above is sufficient for instability, but it is not necessary. We note in passing that \emph{positivity} of the scalar curvature at infinity is neither necessary nor sufficient to avoid Seiberg-Witten instability: see \cite{kn:74} for a discussion.]

While the scalar curvature criterion fails us here, it is however very useful when we \emph{distort} the KMV$_0$ spacetime. If we do so, then the Euclidean version of the spacetime, and likewise its boundary [with metric given in (\ref{P})], will also be distorted. The new metric on the boundary can still be conformally transformed to have constant scalar curvature, but, generically, this new constant will not vanish. Let us see what happens in this case.

There is a large and extremely deep body of mathematical results on the problem of \emph{prescribed scalar curvatures} for compact manifolds. The basic result in the subject is the Kazdan-Warner Classification Theorem \cite{kn:kazdan}, which classifies all compact manifolds into three classes, according to the kinds of scalar functions which can be the scalar curvature of some Riemannian metric on that manifold. We shall not describe this result in detail [see \cite{kn:arrow} for an introduction]; we merely note that deep theorems of Schoen, Yau \cite{kn:schoenyau}, Gromov, and Lawson \cite{kn:lawson} imply that all tori are in ``Kazdan-Warner class Z". This means that the only functions on a torus which can be scalar curvatures are those which are either negative at some point, or identically zero. This implies the following crucial corollary:

\bigskip
\noindent \textsf{COROLLARY (of the theorems of Schoen-Yau-Gromov-Lawson)}:
A generic metric on a torus, when conformally transformed to constant scalar curvature, then has negative scalar curvature.
\bigskip

``Generic" here has a very precise meaning: it means that the scalar curvature of the conformally transformed metric should not be identically zero.

Combining this with the Seiberg-Witten criterion and with the fact that, as we showed in the preceding Section, the Euclidean spaces corresponding to any continuous deformation of the KMV$_0$ metric have toral topology at the boundary, we see that a generic deformation of this black hole will produce a system which is still unstable, in just the same way as the KMV$_0$ spacetime itself. This means that we cannot hope to avoid the instability simply by distorting the boundary geometry in an arbitrary way.

While this is very remarkable, we are not really interested in ``arbitrary" deformations: we want a deformation of a particular type, one that renders the KMV$_0$ velocity profile more realistic. And indeed there is still a loophole: what of non-generic deformations of the KMV$_0$ metric, those which \emph{maintain} zero [Euclidean] scalar curvature at infinity? In particular, can a realistic velocity distribution, like the one pictured in Figure 3, arise from a black hole with scalar curvature at infinity that can be conformally transformed to zero? We now turn to this special but crucial case.

\addtocounter{section}{1}
\section* {\large{\textsf{6. Seiberg-Witten Instability of Deformed KMV$_0$: Scalar-Flat Case}}}
The theorems invoked in the preceding Section have greatly reduced the range of possible black hole deformations which might lead to a stable black hole with a reasonably realistic velocity profile at infinity. But the range of possibilities still seems to be prohibitively large, in that vanishing scalar curvature is [usually] a very weak constraint. For example, there are such metrics on any three-dimensional manifold with the topology of a sphere, any quotient of a sphere,
S$^1$$\times$S$^2$, and any connected sum of these, as well as, of course, on the three-torus and its non-singular quotients. However, the situation in this last case is dramatically different from the others: a theorem due to Bourguignon and later greatly generalised by Gromov and Lawson [see \cite{kn:lawson}, page 306] implies the following extraordinary result\footnote{For simplicity we state the theorem only for tori and their quotients. The theorem is actually far more general, applying to all spin manifolds which are ``enlargeable": see \cite{kn:lawson} for this concept.}.

\bigskip
\noindent \textsf{THEOREM (Bourguignon-Gromov-Lawson)}:
Let g be a Riemannian metric on a torus or on any
non-singular quotient of a torus. If the scalar curvature of
g vanishes, then g is exactly flat, that is, the full curvature tensor is zero.
\bigskip

This enormously simplifies our problem: we now see that the only metrics on the torus which can escape the conclusions of the previous Section are those which are conformally flat.

Conformal flatness for a metric g on a three-dimensional manifold is determined \cite{kn:cotton} by the vanishing of the [symmetric, traceless] \emph{Cotton tensor}
\begin{equation}\label{Q}
\m{C(g)= \eta^{mn(i}\nabla_mR_n^{j)}\,\partial_i\otimes \partial_j,}
\end{equation}
where $\m{\eta^{ijk}}$ is the alternating tensor and R$_{\m{ij}}$ is the Ricci tensor. Requiring this object to vanish imposes third-order differential equations on the metric, so in general we still have a three-parameter family of metrics under this condition. As we shall see, the Euclidean KMV$_0$ boundary metric given in equation (\ref{P}) belongs to this family. We need to determine whether a deformation of this metric, terminating within the conformally flat family, can lead to a Lorentzian black hole metric with a boundary metric generating a velocity profile resembling the one in Figure 3.

We therefore begin with a Euclidean metric on the three-torus, of the form
\begin{equation}\label{R}
\m{g^E_{PC}\; = \; d\tau^2 \;-\;  2\,\Omega(\xi)d\tau d\zeta \;+\; d\xi^2 \;+\;  d\zeta^2,}
\end{equation}
thinking of it as the Euclidean version of the ``peripheral collision metric" given in equation (\ref{B}); see also equation (\ref{P}). [Here, $\Omega(\xi)$ is some periodic function of $\xi$.] The Cotton tensor of this metric, computed using MAPLE, is
\begin{equation}\label{S}
\m{C(g^E_{PC})\; = \;{1\over 2\,(\Omega^2 - 1)^3}\Big[\Omega'''(\Omega^2 - 1)^2\;-\;4(\Omega^2 - 1)\Omega\Omega'\Omega''\;+\;(3\Omega^2 + 1)(\Omega')^3\Big]\Big(- \partial_{\tau}^2\;+\;\partial_{\zeta}^2\Big),}
\end{equation}
where the dash denotes a derivative with respect to $\xi$. Setting this equal to zero will give us the family of Euclidean metrics we seek. It will be convenient to transform the resulting equation to the Lorentzian domain, complexifying $\tau$ and $\Omega$; we denote the complexified version of the latter by $\omega$L [so as to match the notation in equation (\ref{B})]. The result\footnote{Some care is required here. $\Omega$ has to be complexified because of the cross term in equation (\ref{R}); but note that if $\Omega$ is replaced by i$\omega$L in (\ref{S}) then $\Omega'''$ becomes i$\omega'''$L but $\Omega\Omega'\Omega''$ becomes $-$i$\omega\omega'\omega''$L$^3$ and so on, producing a relative minus sign between terms under complexification.} is the equation
\begin{equation}\label{T}
\m{\omega'''(\omega^2L^2 + 1)^2\;-\;4L^2(\omega^2L^2 + 1)\omega\omega'\omega''\;+\;L^2(3\omega^2L^2 - 1)(\omega')^3 \;=\;0,}
\end{equation}
where $\omega$ is now to be regarded as a function of the physical coordinate x, as in Figures 2 and 3, and the dash now denotes a derivative with respect to x. This is a strongly non-linear third-order equation. In order to arrive at definite conclusions, we need to be able to produce graphs of all possible solutions of this equation; surprisingly, that can be done.

As x does not appear explicitly here, this equation is reducible in a standard way to a second-order equation. First we note that $\omega \equiv$ constant is obviously a solution of this equation; it corresponds to a boundary on which the curvature tensor is already zero. Let us set that possibility to one side for the moment ---$\,$ we will return to it below ---$\,$ and define G($\omega$) = (1/2)($\omega')^2$, regarding it as a function of $\omega$. Substituting this into (\ref{T}), we find that the result is a \emph{linear} second-order equation for G:
\begin{equation}\label{U}
\m{(\omega^2L^2 + 1)^2{d^2G \over d\omega^2}\;-\;4L^2(\omega^2L^2 + 1)\omega {dG\over d\omega}\;+\;2L^2(3\omega^2L^2 - 1)G \;=\;0.}
\end{equation}
Still more satisfactory is the fact that this equation can be solved by means of a simple change of variable: setting H = G/(1 + $\omega^2$L$^2$) we find that the equation collapses to d$^2$H/d$\omega^2$ = 0. Thus equation (\ref{T}) reduces to a separable first-order equation:
\begin{equation}\label{V}
\m{{1\over 2}(\omega')^2L^4\;=\;(A\,+\,B\omega L)(1\,+\,\omega^2L^2),}
\end{equation}
where A and B are arbitrary dimensionless constants. [In fact, since $\omega$ should vanish at x = 0 [equation (\ref{C})], we see that A = (1/2)$\omega'(0)^2$L$^4$, so A is either zero or positive.] This equation can of course be solved exactly, but the resulting integrals are not informative. Instead we proceed as follows.

If A and B are both positive, then it is clear that $\omega$ [which should not be negative] is a monotonically increasing function, with a slope which constantly increases, so that we will obtain a graph which never flattens out; in fact it closely resembles the one in Figure 2, not the one in Figure 3. This is not what we want.

If A = 0, then B must be positive, and in this case we obtain a familiar example of an $\omega$ function: since sinl(x) satisfies $\m{({d\over dx}\,sinl(x))^2 = 1 - sinl^4(x)}$, and $\gamma$ = (1 + i)/$\sqrt{2}$ satisfies $\gamma^4 = -1$, it follows that the function sinl$\m{^2\Big(\sqrt{B\over 2}\,\gamma\,{x\over L} + C\Big)/(\gamma^2L}$) satisfies equation (\ref{V}) with A = 0, for any constant C. The latter has to vanish because of condition (\ref{C}), and so we obtain precisely the $\omega$ function [with a particular interpretation of the constant B] which occurs in the boundary metric [equation (\ref{K})] of the KMV$_0$ spacetime. We see that the KMV$_0$ boundary metric is conformally flat, just like the boundary metric of Carter's AdS-Kerr spacetime. [Thus, the deformation we envisage, from Figure 2 to Figure 3, begins and ends [if it exists] in the conformally flat family of metrics.] Figure 2 itself is precisely a graph of this function, with particular parameter values. Again, of course, this is not what we want.

\begin{figure}[!h]
\centering
\includegraphics[width=0.6\textwidth]{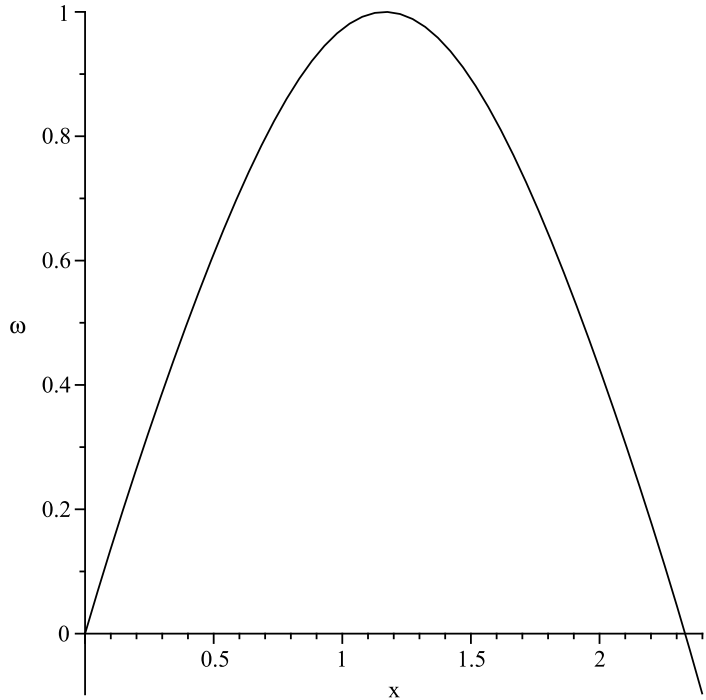}
\caption{Velocity profile, A $>$ 0, B $<$ 0, $\alpha^2 = 1$.}
\end{figure}

The final remaining possibility is that A is positive and B is negative.
Replacing x by x$\m{\sqrt{|B|}}$ and setting $\alpha^2$ = A/$\m{|B|}$, we can write equation (\ref{V}) in the form
\begin{equation}\label{W}
\m{{1\over 2}(\omega')^2\;+\;V(\omega)\;=\;\alpha^2,}
\end{equation}
where for convenience we have set L = 1, and where
\begin{equation}\label{X}
\m{V(\omega)\;=\;\omega^3\;-\;\alpha^2\omega^2\;+\;\omega.}
\end{equation}
The solutions can then be understood by interpreting this as a one-dimensional dynamics problem with x$\m{\sqrt{|B|}}$ playing the role of ``time" and with  ``potential" V($\omega$). Note first that V($\alpha^2$) = $\alpha^2$, which plays the role of the total energy, so the physical region is $0\,\leq\,\omega \, \leq \,\alpha^2$: the $\omega$ function in this case must be bounded, which is highly desirable for our purposes [because an unbounded velocity distribution is precisely what we are striving to avoid]. Furthermore, for $\alpha^2 > \sqrt{3}$, the potential has a local maximum, and we might hope to find solutions such that $\omega$ asymptotically approaches a value which realises that maximum. Unfortunately, if $0\,< \omega \, < \,\alpha^2$ then $\omega^3 - \alpha^2\omega^2 < 0,$ whence it follows that V($\omega$) $<$ $\alpha^2$ for all physical $\omega$: that is, the particle always has more than enough kinetic energy to surmount the maximum, and so it will not stop moving until it reaches $\omega = \alpha^2$. There it will bounce and return to the origin: in other words, $\omega$, as a function of x, increases to a maximum and then decreases back to zero. That is, all $\omega$ functions in this case will resemble the one in Figure 6 [which was produced by means of a numerical solution of (\ref{W}) with $\alpha^2$ = 1]. Again, this is not what we want: while this case respects causality, it entails a \emph{decrease} of the velocity as the boundary of the nucleon is approached, which of course is completely unphysical. [Indeed, this distribution resembles the one induced by a topologically spherical AdS black hole, as discussed in Section 2.]

\begin{figure}[!h]
\centering
\includegraphics[width=0.6\textwidth]{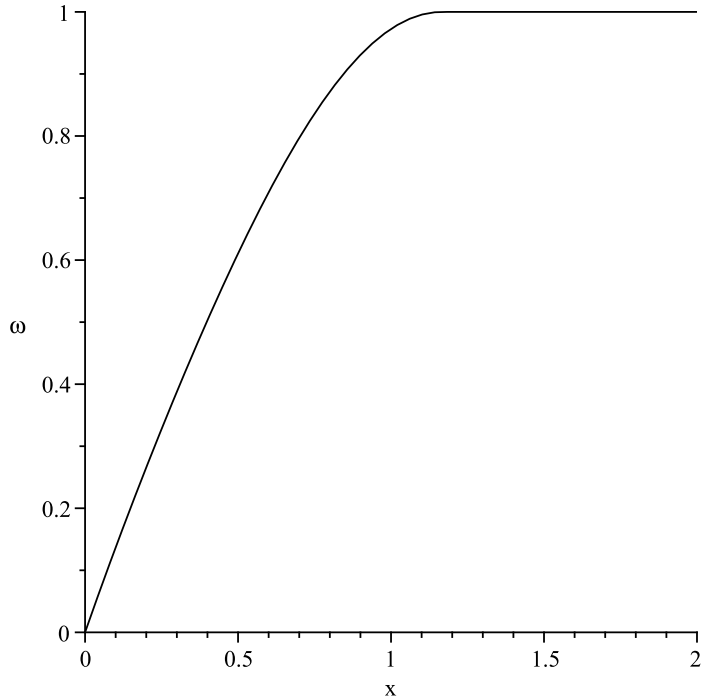}
\caption{Velocity profile, Figure 6 patched with $\omega = 1$.}
\end{figure}

In a final effort to obtain a graph resembling the one in Figure 3, we can try to proceed as follows. As we mentioned earlier, $\omega \equiv$ constant is a solution of equation (\ref{T}); hence we can patch together the increasing part of the graph in Figure 6 with $\omega \equiv \alpha^2 = 1$ to obtain the graph in Figure 7. This might indeed be an acceptable approximate description of a realistic velocity profile; it would correspond to a boundary manifold which is exactly flat in one region but not elsewhere. That is unusual, but such manifolds can in fact be constructed. [They arise in the study of ``local holonomy groups" of Riemannian manifolds \cite{kn:holonomy}.] The problem is that, for reasons explained in Section 3 above, no AdS black hole having such a velocity distribution at infinity can exist: Einstein metrics are analytic in harmonic coordinates, and Figure 7 does not portray an analytic function\footnote{It is interesting to note that the graph in Figure 7 is precisely what MAPLE produces when asked to solve equation (\ref{W}), with $\alpha^2$ = 1, by means of the Runge-Kutta-Fehlberg 4-5th order numerical technique. [Other methods, such as DVERK 7-8th order, simply fail when $\omega$ reaches unity.] The point here is that, when equation (\ref{W}) is solved for $\omega'$, the resulting expression does not satisfy the Lipschitz continuity condition when $\omega = \alpha^2$, which gives rise to difficulties for numerical methods; this is associated with the fact that the Picard existence and uniqueness theorem for first-order ordinary differential equations does not apply at this point. Figure 6 was produced by skipping over the maximum.}. [A change of coordinates could help to raise the level of differentiability ---$\,$ one can readily verify that the coordinates we are using here are not harmonic ---$\,$ but not to that extent.]

In short, a planar AdS black hole with a velocity distribution at infinity similar to the one displayed in Figure 3 cannot induce a conformally flat structure at infinity, simply because all planar AdS black holes with conformally flat boundaries of the correct general form have $\omega$ functions which can be graphed, and none of these graphs remotely resembles Figure 3. A planar AdS black hole with a physical dual geometry must, therefore, induce a scalar curvature at infinity which becomes negative when transformed to a constant. Hence it is unstable, in precisely the same way as a prolate topologically spherical AdS$_5$ black hole with a distortion parameter beyond 2, and in the same way as the KMV$_0$ spacetime: that is, to the uncontrolled nucleation of brane-antibrane pairs.

In fact, while conformal flatness at infinity is necessary for stability against the Seiberg-Witten effect in this situation, it is by no means sufficient, as the example of the KMV$_0$ metric itself shows. We have no reason to believe that any of the geometries discussed in this Section corresponds to a stable black hole, and in fact this seems rather unlikely [particularly in the case where both A and B above are positive, which closely resembles the KMV$_0$ case]. \emph{We conjecture that every rotating planar AdS black hole is unstable in this sense.} Proving this would entail showing that the special black holes we have discussed in this Section are all unstable; as we do not have an explicit construction of them [apart from the KMV$_0$ case, A = 0 above], however, this is still a formidable task.

\addtocounter{section}{1}
\section* {\large{\textsf{7. Conclusion}}}
The principal conclusion of this work is as follows. Consider \emph{any} asymptotically AdS planar black hole with a velocity profile at infinity which increases monotonically away from the central axis, yet remains bounded everywhere, as would certainly be the case in the dual field theory description of the QGP produced in actual collision of heavy ions. Then this black hole is unstable when string theoretic corrections [arising from brane physics] are taken into account. If the AdS/CFT duality correctly describes the dual system in this case, then the implication is that \emph{any quark-gluon plasma produced in a heavy ion collision, with an internal velocity distribution which is even approximately realistic, must be unstable}. This instability was also an important conclusion of \cite{kn:75}, but in that case we used a black hole geometry which was chosen simply because the metric could be written down explicitly. We see now that the instability is \emph{not} an artefact of using that particular geometry. This has been shown by using techniques which, unfortunately, do not allow us to exhibit these other black hole metrics explicitly.

While we have, for convenience, referred to the instability here as an instability of the black hole, this is somewhat misleading. A more accurate description is this: the presence of a rotating planar black hole causes subtle changes in the geometry of the spacetime \emph{far from the black hole}, and it turns out that extended objects like branes, which propagate in that region, are sensitive to these changes, so that their action function becomes negative at all sufficiently large distances from the black hole. In fact, in \cite{kn:75} it was shown that the region in which the brane action becomes negative moves outwards towards infinity as the angular momentum decreases. This implies that the time required for the unstable region to make its presence felt throughout the spacetime becomes longer as the angular momentum is reduced, and we argued that the plasma produced in colliders may be too short-lived for the instability to develop ---$\,$ except possibly for very high values of the ratio of the angular momentum and energy densities [measured in units defined by the critical temperature]. Thus, in practice the instability is not important except for those collisions which involve such high values, and so we spoke of an \emph{angular momentum cutoff}, predicted by holography.

We have seen here that, even if the black hole used in \cite{kn:75} is replaced by one which is dual to a plasma with a realistic distribution of velocity, the system is still unstable; it is therefore safe to assume that the qualitative conclusions we have been discussing continue to hold. In particular, the ``holographic angular momentum cutoff" is a \emph{universal} prediction of holography: it is predicted to be a basic property of \emph{any} dual field theory describing a plasma with a physical velocity distribution and sufficiently large angular momentum density.

A particularly notable finding of \cite{kn:75} was that the time scale of the instability increases surprisingly slowly as the angular momentum decreases. [In particular, it does not diverge as the angular momentum tends to zero.] This is desirable for the application to the QGP, because it is consistent with the fact that the data at relatively low energies do not immediately reveal any cutoff; but it also implies that the time scale is not necessarily very long when the angular momentum is small. This could have important implications for the holography of rotating superconductors. It will be considered elsewhere.

\addtocounter{section}{1}
\section*{\large{\textsf{Acknowledgement}}}
The author is grateful to Prof. Soon Wanmei for support and encouragement.


\begin{thebibliography}{18}
\bibitem{kn:schuk}
J. Schukraft, Results from the first heavy ion run at the LHC,
arXiv:1112.0550 [hep-ex]
\bibitem{kn:schuk2}
Berndt Muller, Jurgen Schukraft, Bolek Wyslouch,
First Results from Pb+Pb collisions at the LHC, arXiv:1202.3233 [hep-ex]
\bibitem{kn:liang}
Zuo-Tang Liang, Xin-Nian Wang, Globally polarized quark-gluon plasma in non-central A+A collisions,
Phys.Rev.Lett. 94 (2005) 102301, Erratum-ibid. 96 (2006) 039901, \x nucl-th/0410079
\bibitem{kn:bec}
F. Becattini, F. Piccinini, J. Rizzo, Angular momentum conservation in heavy ion collisions at very high energy, Phys.Rev.C77:024906,2008,
arXiv:0711.1253 [nucl-th]
\bibitem{kn:huang}
Xu-Guang Huang, Pasi Huovinen, Xin-Nian Wang, Quark Polarization in a Viscous Quark-Gluon Plasma,
Phys. Rev. C84, 054910(2011), arXiv:1108.5649 [nucl-th]
\bibitem{kn:solana}
Jorge Casalderrey-Solana, Hong Liu, David Mateos, Krishna Rajagopal, Urs Achim Wiedemann,
Gauge/String Duality, Hot QCD and Heavy Ion Collisions, arXiv:1101.0618 [hep-th]
\bibitem{kn:pedraza}
Mariano Chernicoff, J. Antonio Garcia, Alberto Guijosa, Juan F. Pedraza,
Holographic Lessons for Quark Dynamics, J.Phys.G G39 (2012) 054002, arXiv:1111.0872 [hep-th]
\bibitem{kn:carter}
B. Carter, Hamilton-Jacobi and Schrodinger separable solutions of Einstein's equations,
Commun.Math.Phys. 10 (1968) 280
\bibitem{kn:hawrot}
S.W. Hawking, C.J. Hunter, Marika Taylor, Rotation and the AdS/CFT correspondence, Phys.Rev.D59:064005,1999, \x hep-th/9811056
\bibitem{kn:schalm}
A. Nata Atmaja, K. Schalm, Anisotropic Drag Force from 4D Kerr-AdS Black Holes, JHEP 1104:070,2011,
arXiv:1012.3800 [hep-th]
\bibitem{kn:75}
Brett McInnes, Fragile Black Holes and an Angular Momentum Cutoff in Peripheral Heavy Ion Collisions, Nucl. Phys. B 861 (2012) 236, arXiv:1201.6443 [hep-th]
\bibitem{kn:confined}
Edward Witten, Anti-de Sitter space, thermal phase transition, and confinement in gauge theories,
Adv.Theor.Math.Phys.2:505,1998, \x hep-th/9803131
\bibitem{kn:klemm}
D. Klemm, V. Moretti, L. Vanzo, Rotating Topological Black Holes, Phys.Rev.D57:6127,1998; Erratum-ibid.D60:109902,1999,
arXiv:gr-qc/9710123
\bibitem{kn:74}
Brett McInnes, Kerr Black Holes Are Not Fragile,
Nucl. Phys. B 857 (2012) 362, arXiv:1108.6234 [hep-th]
\bibitem{kn:seiberg}
Nathan Seiberg, Edward Witten, The D1/D5 System And Singular CFT,
JHEP 9904 (1999) 017, \x hep-th/9903224
\bibitem{kn:bemo}
Victor Roy, A. K. Chaudhuri, Bedangadas Mohanty,
Comparison of results from a 2+1D relativistic viscous hydrodynamic model to elliptic and hexadecapole flow of charged hadrons measured in Au-Au collisions at $\sqrt{s_{\rm {NN}}}$ = 200 GeV,
arXiv:1204.2347 [nucl-th]
\bibitem{kn:subir}
Subir Sachdev, What can gauge-gravity duality teach us about condensed matter physics?,
Annual Review of Condensed Matter Physics 3, 9 (2012), arXiv:1108.1197 [cond-mat.str-el]
\bibitem{kn:sonner}
Julian Sonner, A Rotating Holographic Superconductor, Phys.Rev.D80:084031,2009, arXiv:0903.0627 [hep-th]
\bibitem{kn:leigh}
Robert G. Leigh, Anastasios C. Petkou, P. Marios Petropoulos,
Holographic Three-Dimensional Fluids with Nontrivial Vorticity, Phys.Rev. D85 (2012) 086010, arXiv:1108.1393 [hep-th]
\bibitem{kn:leigh2}
Robert G. Leigh, Anastasios C. Petkou, P. Marios Petropoulos, Holographic Fluids with Vorticity and Analogue Gravity, arXiv:1205.6140 [hep-th]
\bibitem{kn:AdSRN}
Brett McInnes,
Bounding the Temperatures of Black Holes Dual to Strongly Coupled Field Theories on Flat Spacetime, JHEP09(2009)048,
arXiv:0905.1180 [hep-th]
\bibitem{kn:triple}
Brett McInnes, Holography of the Quark Matter Triple Point,
Nucl.Phys. B832 (2010) 323, arXiv:0910.4456 [hep-th]
\bibitem{kn:73}
Brett McInnes, A Universal Lower Bound on the Specific Temperatures of AdS-Reissner-Nordstrom Black Holes with Flat Event Horizons, Nucl.Phys.B848 (2011) 474, \x 1012.4056 [hep-th]
\bibitem{kn:ong1}
Yen Chin Ong, Stringy Stability of Dilaton Black Holes in 5-Dimensional Anti-de Sitter Space, Proceedings of Conference: C10-02-24, p.583-590,
arXiv:1101.5776 [hep-th]
\bibitem{kn:ong2}
Yen Chin Ong, Pisin Chen, Stringy Stability of Charged Dilaton Black Holes with Flat Event Horizons, arXiv:1205.4398 [hep-th]
\bibitem{kn:visser}
Matt Visser, The Kerr spacetime: A brief introduction, arXiv:0706.0622 [gr-qc]
\bibitem{kn:lemmo}
J.P.S. Lemos, Phys.Lett.B353:46,1995,
\x gr-qc/9404041; R.B. Mann, Class.Quant.Grav. 14 (1997) L109, arXiv:gr-qc/9607071;
Rong-Gen Cai, Yuan-Zhong Zhang, Phys.Rev.D54:4891,1996, \x gr-qc/9609065;
Danny Birmingham, Class.Quant.Grav. 16 (1999) 1197, arXiv:hep-th/9808032
\bibitem{kn:murata}
Keiju Murata, Tatsuma Nishioka, Norihiro Tanahashi,
Warped AdS$_5$ Black Holes and Dual CFTs, Prog.Theor.Phys.121:941,2009, arXiv:0901.2574 [hep-th]
\bibitem{kn:kunz}
Yves Brihaye, Jutta Kunz, Eugen Radu,
From black strings to black holes: Nuttier and squashed AdS(5) solutions, JHEP 0908:025,2009, arXiv:0904.1566 [gr-qc]
\bibitem{kn:brihaye}
Yves Brihaye, Eugen Radu, Cristian Stelea, Charged squashed black holes with negative cosmological constant in odd dimensions, Phys.Lett. B709 (2012) 293,
arXiv:1201.3098 [hep-th]
\bibitem{kn:porrati}
M. Kleban, M. Porrati, R. Rabadan, Stability in Asymptotically AdS
Spaces, JHEP 0508 (2005) 016, \x hep-th/0409242
\bibitem{kn:wittenads}
Edward Witten,
Anti-de Sitter space and holography, Adv.Theor.Math.Phys.2:253-291,1998,
\x hep-th/9802150
\bibitem{kn:kengo}
Kengo Maeda, Takashi Okamura, Jun-ichirou Koga,
Inhomogeneous charged black hole solutions in asymptotically anti-de Sitter spacetime, Phys.Rev. D85 (2012) 066003, arXiv:1107.3677 [gr-qc]
\bibitem{kn:chrusc1}
Michael T. Anderson, Piotr T. Chrusciel, Erwann Delay,
Nontrivial, static, geodesically complete, vacuum space-times with a negative cosmological constant, JHEP 0210:063,2002, \x gr-qc/0211006
\bibitem{kn:chrusc2}
M. Anderson, P.T. Chrusciel, E. Delay,
Non-trivial, static, geodesically complete space-times with a negative cosmological constant II. $n\ge 5$,
Proceedings of the Strasbourg Meeting on AdS-CFT correspondence, O.Biquard, V.Turaev Eds., IRMA Lectures in Mathematics and Theoretical Physics, de Gruyter, Berlin, New York, arXiv:gr-qc/0401081
\bibitem{kn:horror}
Oscar J.C. Dias, Gary T. Horowitz, Jorge E. Santos,
Black holes with only one Killing field, JHEP 1107 (2011) 115, arXiv:1105.4167 [hep-th]
\bibitem{kn:fragile}
Brett McInnes,
Fragile Black Holes, Nucl.Phys. B842 (2011) 86, arXiv:1008.0231 [hep-th]
\bibitem{kn:delay}
P.T.Chrusciel, E.Delay, Non-singular, vacuum, stationary space-times with a negative cosmological constant,
Annales Henri Poincare 8: 219,2007,
arXiv:gr-qc/0512110
\bibitem{kn:turk}
Dennis M. DeTurck, Jerry L. Kazdan, Some regularity theorems in Riemannian geometry, Annales Scientifiques de l'École Normale Supérieure. Quatrième Série 14 (3) (1981) 249
\bibitem{kn:abramo} M. Abramowitz and I.A. Stegun, \emph{Handbook of Mathematical
Functions With Formulas, Graphs, and Mathematical Tables}, Dover
Publications, New York, 1964
\bibitem{kn:lemniscate}
Weisstein, Eric W., Lemniscate Function. From MathWorld--A Wolfram Web Resource. http://mathworld.wolfram.com/LemniscateFunction.html
\bibitem{kn:lemnconst}
Weisstein, Eric W., Lemniscate Constant. From MathWorld--A Wolfram Web Resource. http://mathworld.wolfram.com/LemniscateConstant.html
\bibitem{kn:schoenyam}
R. Schoen, Conformal deformation of a Riemannian metric to constant
scalar curvature. J. Differential Geom. 20 (1984) 479
\bibitem{kn:kazdan}
Kazdan, Jerry L., Warner, F. W., Existence and conformal
deformation of metrics with prescribed Gaussian and scalar
curvatures. Ann. of Math. 101 (1975) 317
\bibitem{kn:arrow}
Brett McInnes, Arrow of Time in String Theory, Nuclear Physics B782
(2007) 1, \x hep-th/0611088
\bibitem{kn:schoenyau}
R. Schoen, S.-T. Yau, Existence of incompressible minimal surfaces
and the topology of three-dimensional manifolds with nonnegative
scalar curvature, Ann. of Math. 110 (1979) 127
\bibitem{kn:lawson}
H. Blaine Lawson and Marie-Louise Michelsohn, \emph{Spin Geometry},
Princeton University Press, 1990
\bibitem{kn:cotton}
A. Garcia, F.W. Hehl, C. Heinicke, A. Macias,
The Cotton tensor in Riemannian spacetimes, Class.Quant.Grav. 21 (2004) 1099,
arXiv:gr-qc/0309008
\bibitem{kn:holonomy}
Brett McInnes, Obtaining holonomy from curvature, J. Phys. A: Math. Gen. 30 (1997) 661


\end{thebibliography}
\end{document}